\documentclass[]{raa}            
\usepackage{graphicx,times}
\usepackage{natbib}

\providecommand{\bm}[1]{\mbox{\boldmath$#1$\unboldmath}}

\begin{document}

   \title{ The properties of the high-mass star
   formation region IRAS 22475+5939
  }

 \volnopage {{\bf 20xx} Vol.\ {\bf x} No. {\bf XX}, 000--000}
   \setcounter{page}{1}

   \author{X.-L. Liu
   \inst{1,2}
   \and J.-J. Wang
   \inst{1}
}

   \institute{{National Astronomical Observatories, Chinese Academy of Sciences,
             Beijing 100012, China; {\it liuxiaolan10@mails.gucas.ac.cn} \\
             \and
                  Graduate University of the Chinese Academy of
                  Sciences, Beijing, 100080, China \\
}
 \vs \no
   {\small 26 March 2012 }
}

\abstract{{ IRAS22475+5939 has been well researched by previous
astronomers. But we still get some new characteristics about it,
using the first observations in lines of $\rm CO
\,J=2-1$,${}^{13}\rm CO \,J=2-1$,${}^{13}\rm \,CO J=3-2$ by the
KOSMA 3 m telescope. The mapping of the intensity ratio of,Eiroa
1981,Felli81,Eiroa 1981 ${}^{13}\rm CO\, J=3-2  $ and ${}^{13}\rm CO
\,J=2-1$ shows the distribution of the temperature with two peaks,
which don't coincide with IRAS22475+5939 source and the center of
the HII region, but at the edge of the HII region. The overlays of
the Spitzer IRAC $8\,\mu m $ and CO contours indicate that they are
associated with each other and the strongest polycyclic aromatic
hydrocarbons (PAHs) emission is at the position of IRAS22475+5939
source. While the IRAS LRS spectrum at $7\,\mu m \sim 23\,\mu m  $
and the PHT-s spectrum at $2\,\mu m \sim 12\,\mu m $ of
IRAS22475+5939 source also exhibit strong PAHs emission characters
at the main PAH bands. The diversity of PAH family should be
responsible for the plateaus of PAHs emission in the PHT-s spectrum
and the IRAS-LRS spectrum. An analysis and modeling in infrared
bands suggest that IRAS22475+5939 is more likely to be a Class I
YSO. Where this is the case, the star is likely to have a
temperature $\rm T_{EFF} \,\sim 9995.8 \,K , mass \,\sim 15.34\,
M_\odot, lumilosity\, \sim 1.54\times 10^4 \,L_\odot,\, and
\,age\,\sim 1.54\times 10^4\, yr $. The model shows that the
circumstellar disc emission is important for the wavelength between
$1 \rm \,and \,10 \,\mu m$, otherwise, envelope fluxes for $\lambda
>10 \mu m $. The bipolar outflow is confirmed in the molecular
cloud. The excited star of the HII region has the chance to be the
driving source of the outflow. The high resolution is required.
 }
    }
   \authorrunning{X.-L. Liu \& J.-J Wang}
   \titlerunning{ The properties in  IRAS 22475+5939 source}

   \maketitle

\section{Introduction}
\label{sect:intro} Researches of massive star formation have been
getting significant attention in the recent years. But the mechanism
of how massive stars form is still debatable, whether it is through
the process similar to that for the low mass stars, ie., through
disk accretion and driving molecular outflows (Shu et al. 1987), or
through alternative scenarios-coalescence of low mass stars (Bonnell
et al. 1998). But the recent observations show that outflows are
common in the massive star formation regions (Zhang et al. 2001) and
it is almost clear that stars at least up to late-O spectral types
form primarily through disk accretion (Varricatt 2012). HII regions
will form when the massive stars ionize their surroundings and
provide information about massive star formation within molecular
clouds (Heyer et al. 1989;Churchwell 2002). Therefore, the forming
stars will have a big influence on their surrounding environment
during the evolutional processes. It is an indirect way to study the
properties in the star formation regions in order to know more about
how massive stars form.

IRAS22475+5939 has been detected in the sharpness HII region S146
 (Zuckerman \& Evans 1974). The distance of IRAS22475+5939
from us is about 4.7 kpc  (Henkel 1986). Eiroa
 (1981) believed S146 was excited by an IRS 1, locating at position $\alpha(B1950)=22^{\rm
h}47^{\rm m}29^{\rm s}.7$, $\delta(B1950)=+59\dg 38'55''\pm 7 $ at
$2.2\,\mu m $, $\alpha(B1950)=22^{\rm h}47^{\rm m}29^{\rm s}.5$,
$\delta(B1950)=+59\dg 39'01''\pm 3$\, at $0.9\,\mu m $. Blair (1978)
found a $\rm H_2O$ maser associated with IRAS22475+5939. A 2MASS
source 22492900+59545600( $\alpha=22^{\rm h}49^{\rm m}29^{\rm
s}.06$, $\delta=+59\dg 54'55''$.7, $\rm J2000$ ) was detected by
Wang(1997). Felli (1981) showed a bipolar outflow with the optical
and $6\,cm $ emissions. The bipolar molecular outflow was also
confirmed from surveys of the $\rm CO\,J=1-0$ (Yang \& Wu 1998),
$\rm CO \,J=2-1$ (Jiang et al. 2001 ) and $\rm CO \,J=3-2$ (Wu et
al. 2005) and Jiang (2001) thought a massive star was forming in the
molecular cloud. Guan (2008) made a mapping survey of the massive CO
cores with the $\rm CO,\,{}^{13}CO ,\,{}^{18}CO \,J=1-0$ lines. The
contour of $\rm {}^{13}CO\,J=1-0$ was studied to be associated with
MSX $8\,\mu m$ emission (Guan \& Wu 2008). The IRAS-LRS (Low
Resolution Spectra) spectrum was analyzed by Muizon (1990), Volk
(1991) and Chen (1995), respectively. The LRS spectrum shows strong
[NeII]($12.8\,\mu m$), [NeIII]($15.5\,\mu m$), [SIII]($18.7\,\mu
m$), PAHs emission ($7.7\,\mu m, 8.6\,\mu m, 11.3\,\mu m$) as well
as the sillicate absorption feature at $9.7\,\mu m$ (Muizon et al.
1990; Volk \& Kwok 1991; Chen et al. 1995).

In this paper, we show the existence of the outflow and discuss the
properties in the observed region.  ${}^{13}\rm CO $ is optically
thin and can trace the internal region of the molecular core. And
the intensity ratio $\rm R_{I_{32/21}}$ (ratio of ${}^{13}\rm CO \,
J=3-2$ and  ${}^{13}\rm CO \,J=2-1)$ contains the information about
the temperature distribution in the molecular cloud, and
mid-infrared emission at $8\,um $ is thought to be from small dust
grains and polycyclic aromatic hydrocarbons (PAHs), and excited by
the UV radiation leaking from the HII regions (Leger \& Puget 1984;
Deharveng et al. 2003,2005). We also use the Stokes I image from the
observations for the $1.4$ GHZ NRAO VLA Sky Survey(NVSS) to trace
the HII region in order to know the relationship between the
molecular cloud, the intensity ratio and the HII region. We describe
the observation in section 2. The results and discussion of the
properties are in section 3. section 4 makes a conclusion.

\section{Observations and data reduction}
\label{sect:Obs}

\subsection{The observational data}
We carried out observations toward IRAS22475+5939 ( $\alpha=22^{\rm
h}49^{\rm m}29^{\rm s}.4$, $\delta=+59\dg 54'54''$.00, $\rm J2000$)
source in $\rm CO\, J=2-1$,${}^{13}\rm CO \,J=2-1$,${}^{13}\rm CO
 \,J=3-2 $  lines using the KOSMA 3 m telescope at Gornergrat,
Switzerland in April 2004. The half-power beam widths of the
telescope at observing frequencies $230.538$ GHZ, $220.399$ GHZ,
$330.588$ GHZ are $130''$, $130''$ and $80''$ , respectively. The
pointing and tracking accuracy is better than $10''$. The DSB
receiver noise temperature is about $120$ K. The medium and variable
resolution acousto-optical spectrometers have 1501 and 1601
channels. The channel widths of $248$ MHZ and $544$ MHZ correspond
to velocity resolutions of $0.21\, \rm km\cdot s^{-1}$ and $0.29 \,
\rm km\cdot s^{-1} $, respectively. The beam efficiency is $0.68 $
at $230$ GHZ and $220$ GHZ, but $0.72$ at $330$ GHZ and $345$ GHZ.
The forward efficiency is $0.93$. The $ 80''$ resolution of the
$J=3-2$ data was convolved to $130''$ with an effective beam size
$\sqrt{(130^2-80^2)}=102''$. The correction for the line intensities
to the main beam temperature scale was made using the formula $\rm
T_{mb}=\rm(F_{eff}/B_{eff}\ast T_A^*)$. The data were reduced by the
software CLASS (Continuum and Line Analysis Single-Disk Software)
and GREG (Grenoble Graphic).
\subsection{Archival data}
\begin{figure}
     \begin{minipage}[t]{0.495\linewidth}
  \centering
   \includegraphics[width=60mm,height=55mm,angle=0]{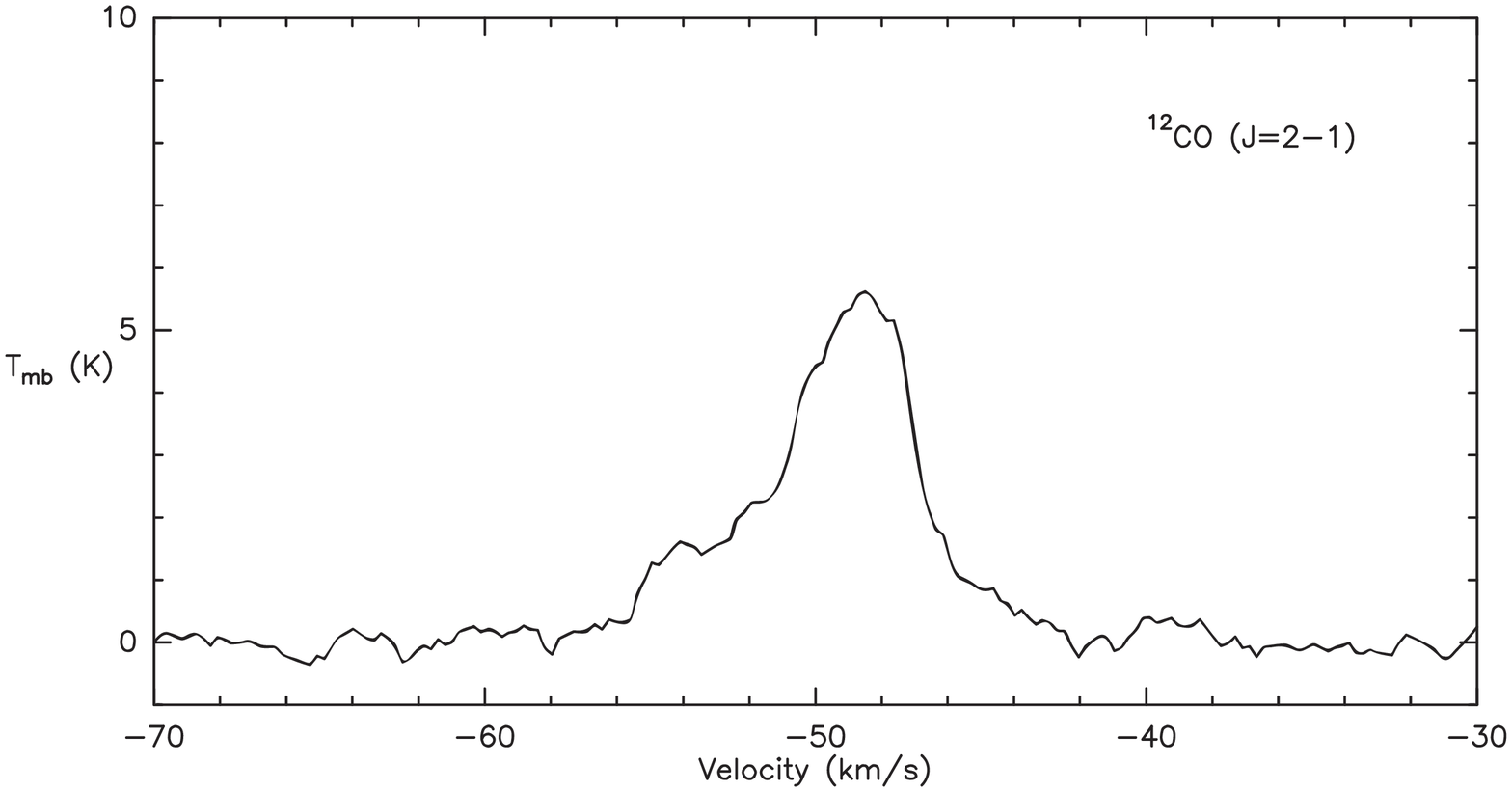}
  \end{minipage}%
  \begin{minipage}[t]{0.495\textwidth}
  \centering
   \includegraphics[width=50mm,height=40mm,angle=0]{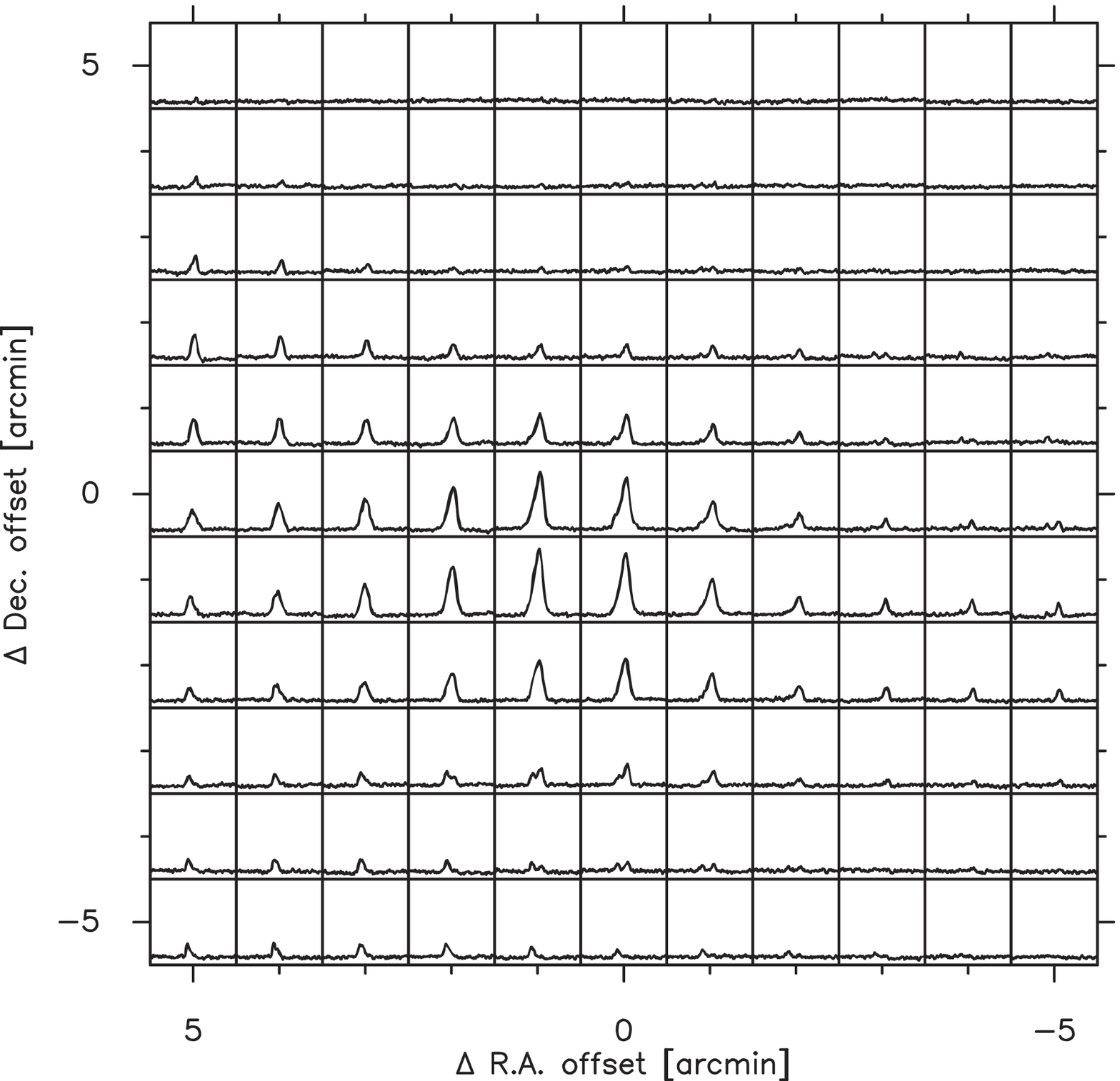}
  \end{minipage}%
\end{figure}
\begin{figure}
   \begin{minipage}[t]{0.495\linewidth}
  \centering
   \includegraphics[width=60mm,height=35mm,angle=0]{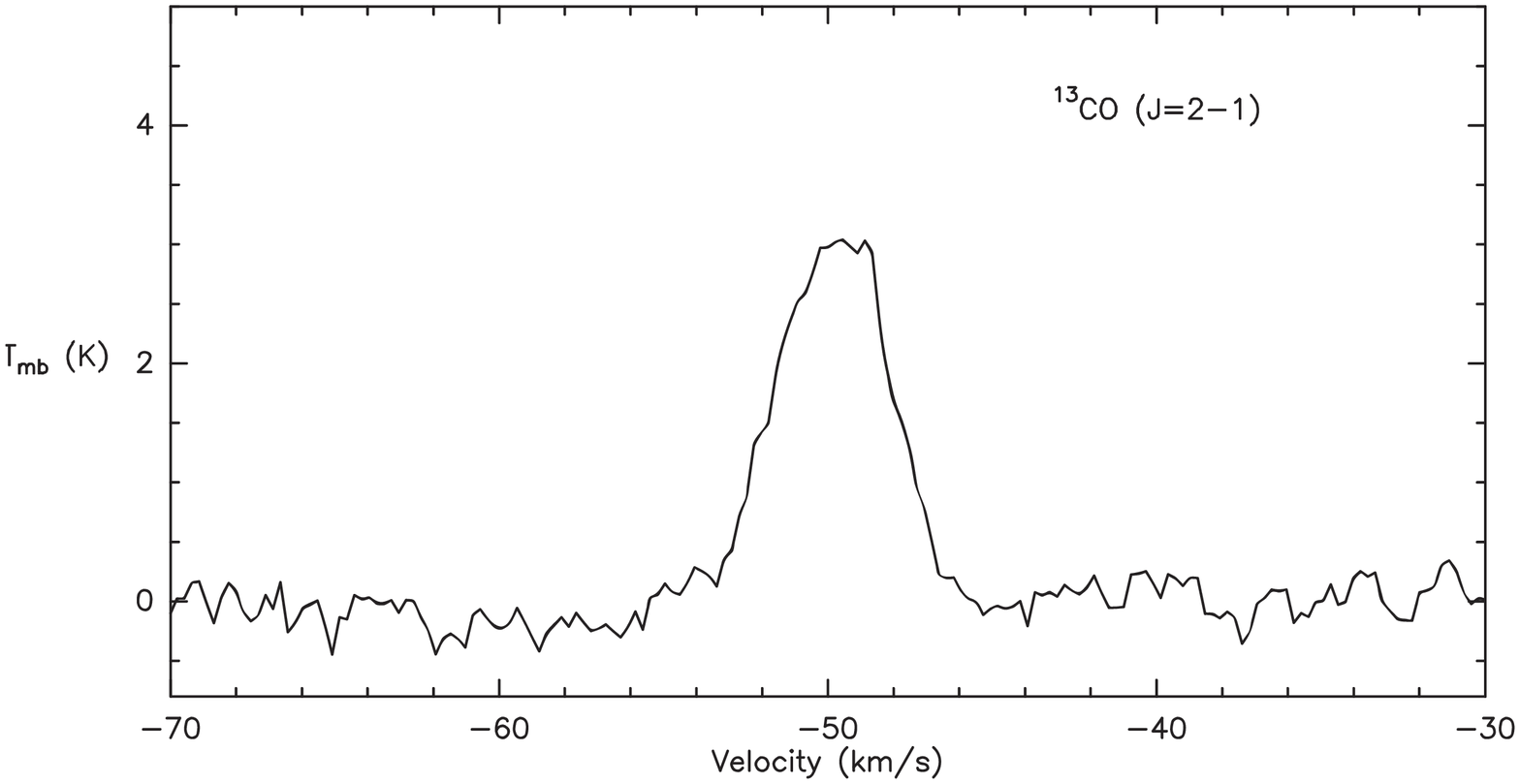}
  \end{minipage}%
  \begin{minipage}[t]{0.495\textwidth}
 \centering
   \includegraphics[width=50mm,height=40mm,angle=0]{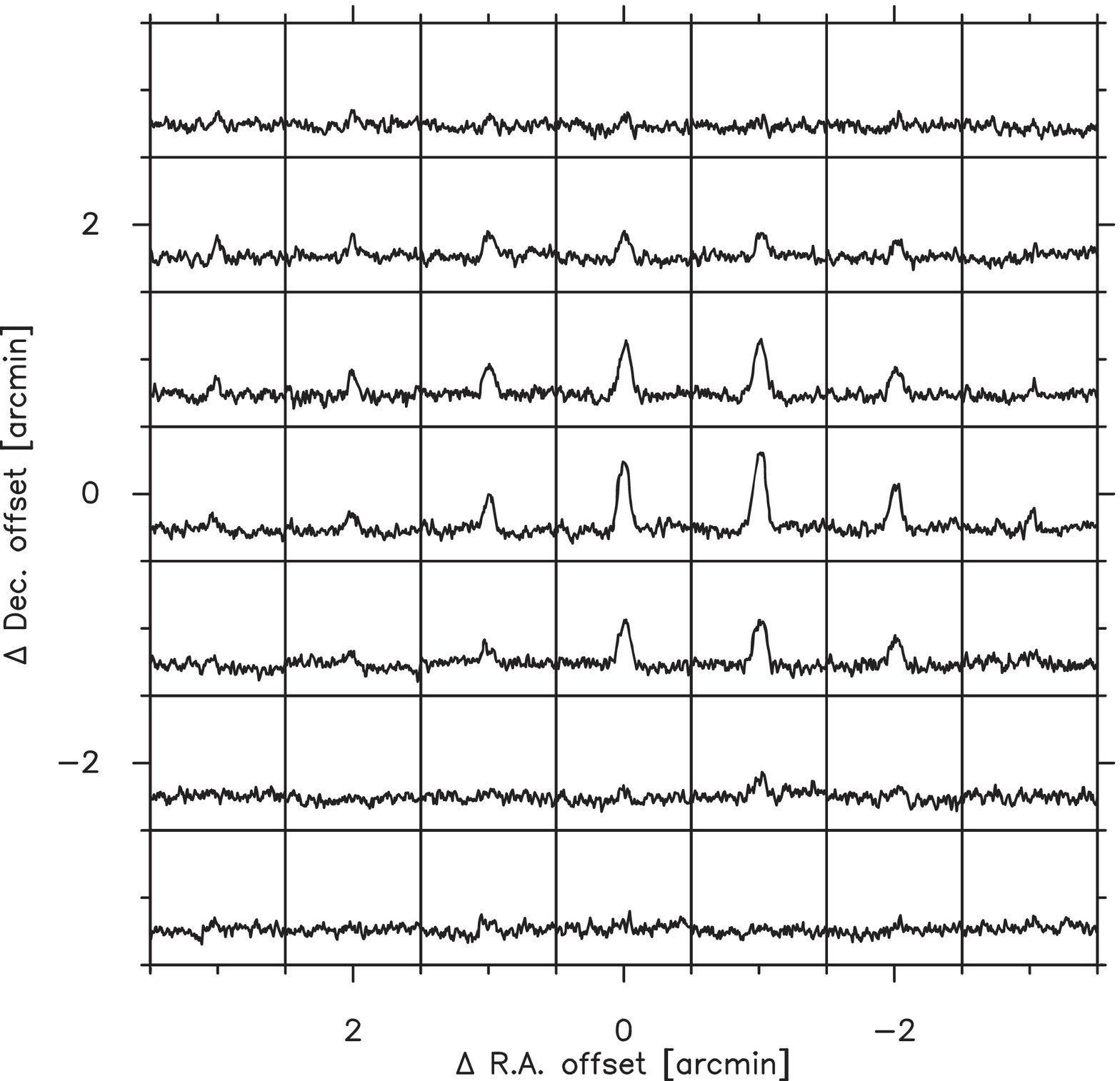}
  \end{minipage}%
\end{figure}
\begin{figure}
   \begin{minipage}[t]{0.495\linewidth}
  \centering
   \includegraphics[width=60mm,height=35mm,angle=0]{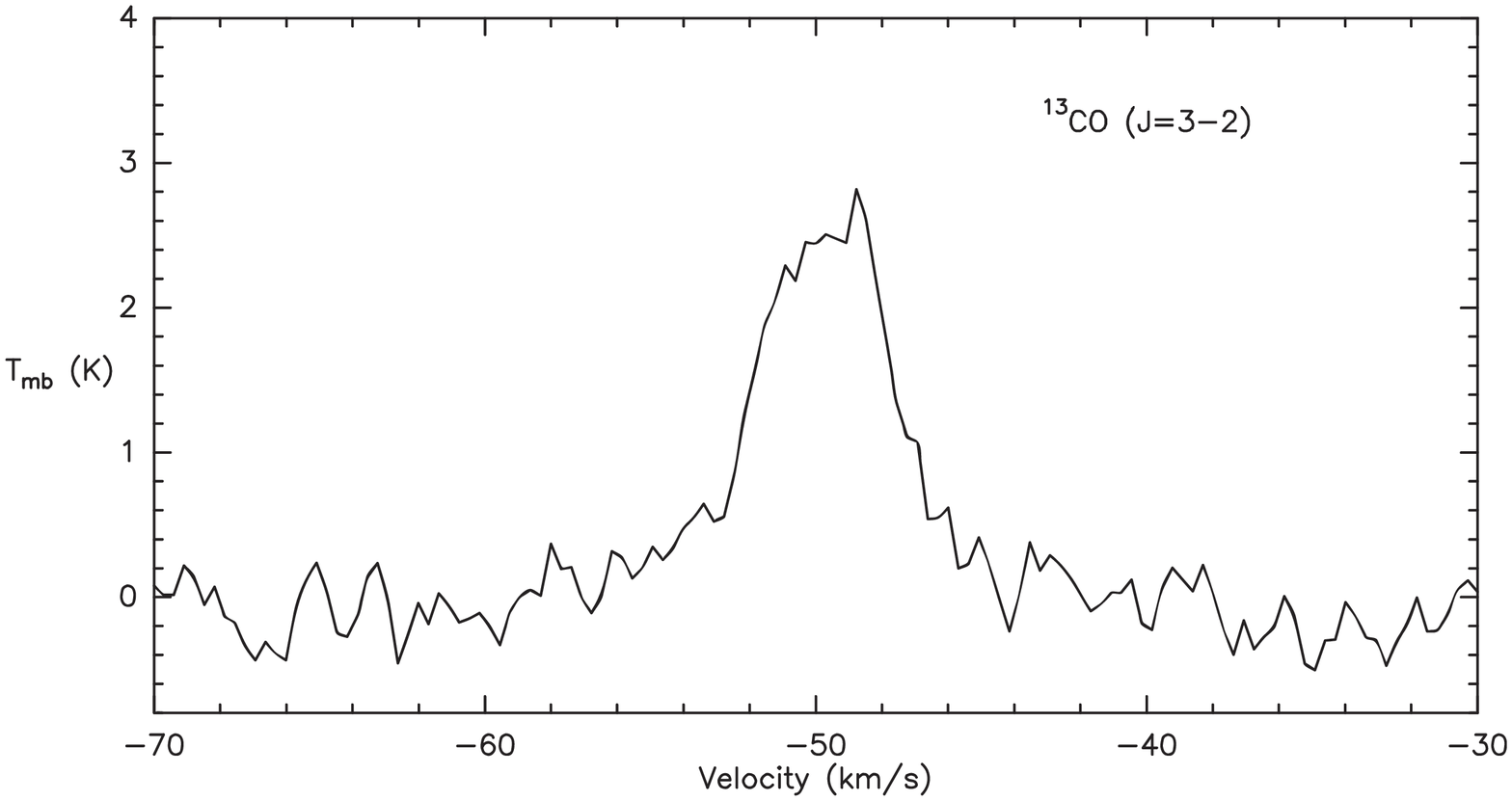}
  \end{minipage}%
  \begin{minipage}[t]{0.495\textwidth}
  \centering
   \includegraphics[width=50mm,height=40mm,angle=0]{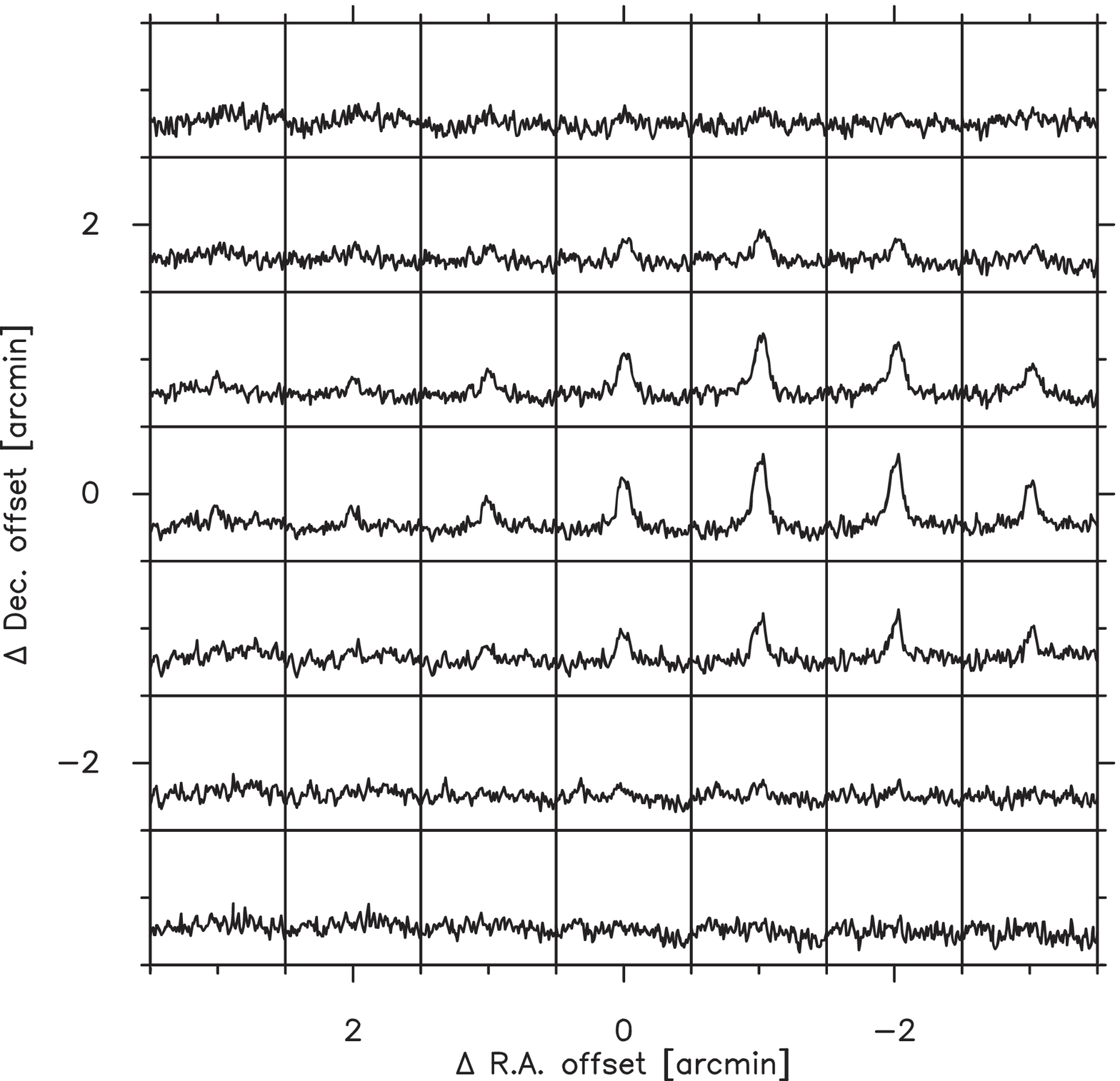}
  \end{minipage}%
   \caption{Left: spectra of $\rm CO\, (J=2-1)$, ${}^{13}\rm CO \,(J=2-1)$ and
   ${}^{13}\rm CO\, (J=3-2)$ at the position of (-1, 0).
    Right: channel maps of the corresponding isotopes
   }
   \label{Fig 1}
\end{figure}
The Spitzer IRAC $8\,um$ imaging is available toward the direction
of IRAS22475+5939 at an angular resolution of $\sim 2''$. S146 was
detected well by the $1.4$ GHZ NRAO VLA Sky Survey(NVSS). Its
resolution is $45''$ and a limiting peak source brightness is about
$2.5$ mJy/beam. We have compiled the SED for IRAS22475+5939 source
using near-IR (JHK) flux from the 2MASS All-Sky Point Source Catalog
(PSC), mid-IR MSX four bands from the Midcourse Space Experiment
(MSXC6), and the IRAS data from IRAS Point Source Catalog v2.1
(PSC). The IRAS LRS spectrum is obtained from the LRS database at
the University of Calgary\footnote{See
http://www.astro.wisc.edu/protostars} (Hodge et al. 2004;Kwok et
al.1997). The LRS was a slitless spectrometer and well-suited to
point sources whereas confusion arises in the case of extended
source. Two wavelength channels are recorded simultaneously: Band 1
($7.7-13.4 \,\mu m$) and Band 2 ($11.0-22.6 \,\mu m$). The
wavelength resolution $\lambda /\Delta \lambda$ varies between 10
and 60; it increases with wavelength inside each wavelength band,
the resolution being systematically somewhat lower in band 2 than in
band 1. PHT-s consist of a dual grating spectrometer with two
64-element arrays that span the $2.5-4.9 \,\mu m$ and $5.8-11.6
\,\mu m$ spectra regions. The Auto Analysis Result produced by the
Off-Line Processing(OLP) from the ISO Data Archive(IDA) is available
(Hodge et al. 2004;Lemke et al.1996)

\section{Results and discussion}
\label{sect:Res}

\subsection{The outflow }
\subsubsection{The molecular spectral lines}

Figure 1 shows the spectra and channel maps of $\rm CO \,J=2-1$,
${}^{13}\rm CO\, J=2-1$, ${}^{13}\rm CO\, J=3-2$. The spectra are at
the positions (-1,0). Each spectrum shows a broad width as well as
the almost same $\rm V_{SLR}$ and shape. The showing spectra are the
strongest for ${}^{13}\rm CO\, J=2-1$, ${}^{13}\rm CO\, J=3-2$. The
optically thick line $\rm CO \,J=2-1$
 is not gaussian shape and has wings. The blue wing and the red wing
are not asymmetrical. Table 1 presents the observational parameters
of IRAS22475+5939 source. The ranges of the FW come from P-V
diagram, which shows the high-velocity gas exists.

\begin{table}[h]
\bc
 \small
 \centering

\begin{minipage}[]{100mm}

 \caption[]{Observation physical parameters of IRAS22475+5939 Source}
 \label{Tab 1} \end{minipage}
\begin{tabular}{ccccc}
 \hline\noalign{\smallskip}
$\rm Name$        & $\rm T_{mb}$  &$\rm FWHM $          &$\rm FW $                &$\rm V_{LSR}$  \\
             & $\rm (K)$   &$\rm(km\cdot s^{-1})$&$\rm (km\cdot s^{-1})$    &$\rm (km\cdot s^{-1})$\\  \hline\noalign{\smallskip}
$\rm CO\,(J=2-1)$    & $11.20$ & $4.99\pm0.03$   & $12.37$             & $-49.60\pm0.01$  \\
${}^{13}\rm CO\,(J=2-1)$ & $3.02$  & $3.90\pm0.08$   & $6.33 $             & $-49.80\pm0.04 $  \\
${}^{13}\rm CO\,(J=3-2)$ & $2.55 $ & $4.79\pm 0.17$  & $7.87$              & $-49.69\pm0.07$   \\
     \noalign{\smallskip}\hline
       \end{tabular}
       \ec
       \end{table}

\subsubsection{The outflow and its physical parameters}
\begin{figure}
  \begin{minipage}[t]{0.495\linewidth}
  \centering
   \includegraphics[width=60mm,height=55mm]{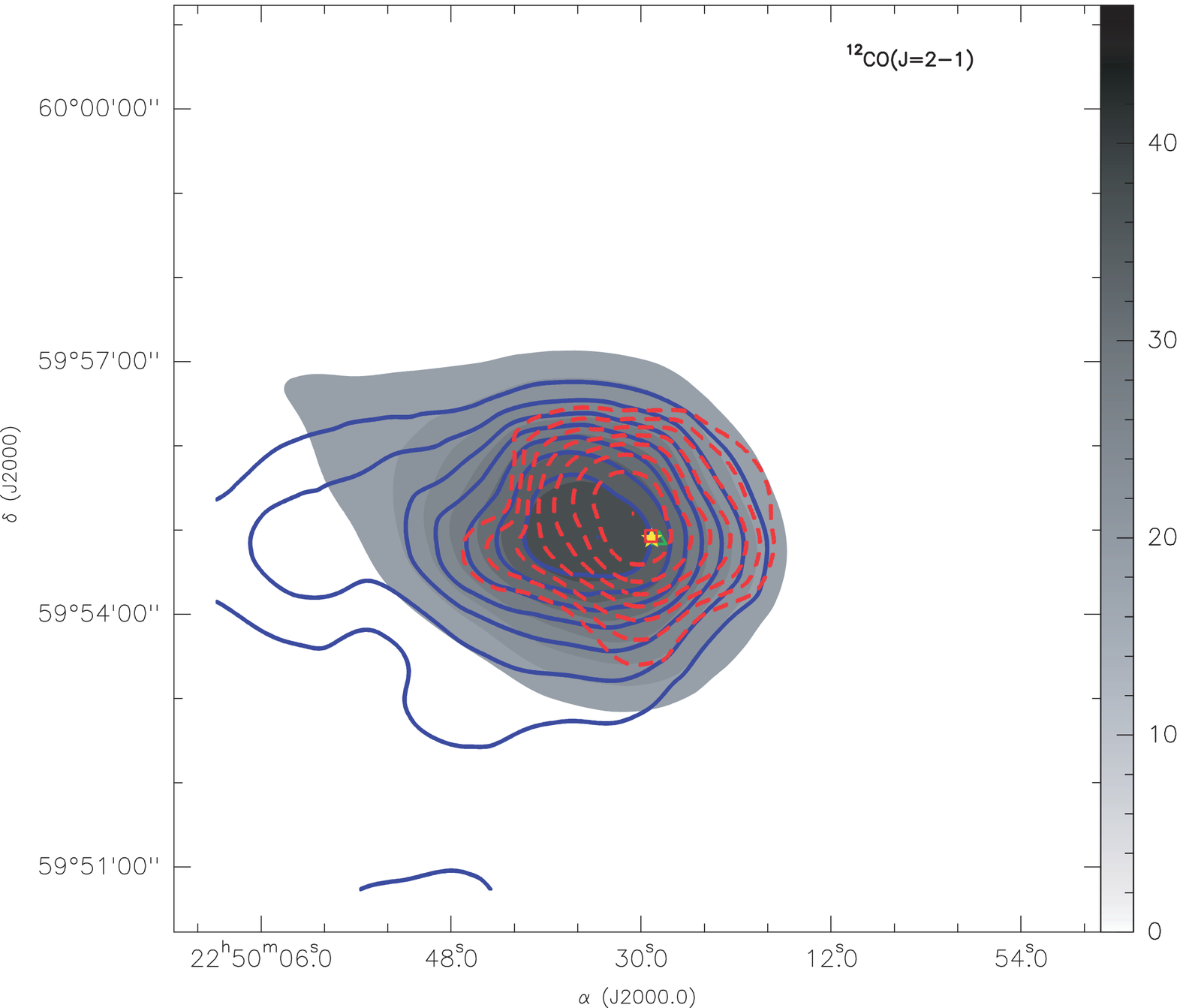}
  \end{minipage}%
  \begin{minipage}[t]{0.495\textwidth}
  \centering
   \includegraphics[width=60mm,height=55mm]{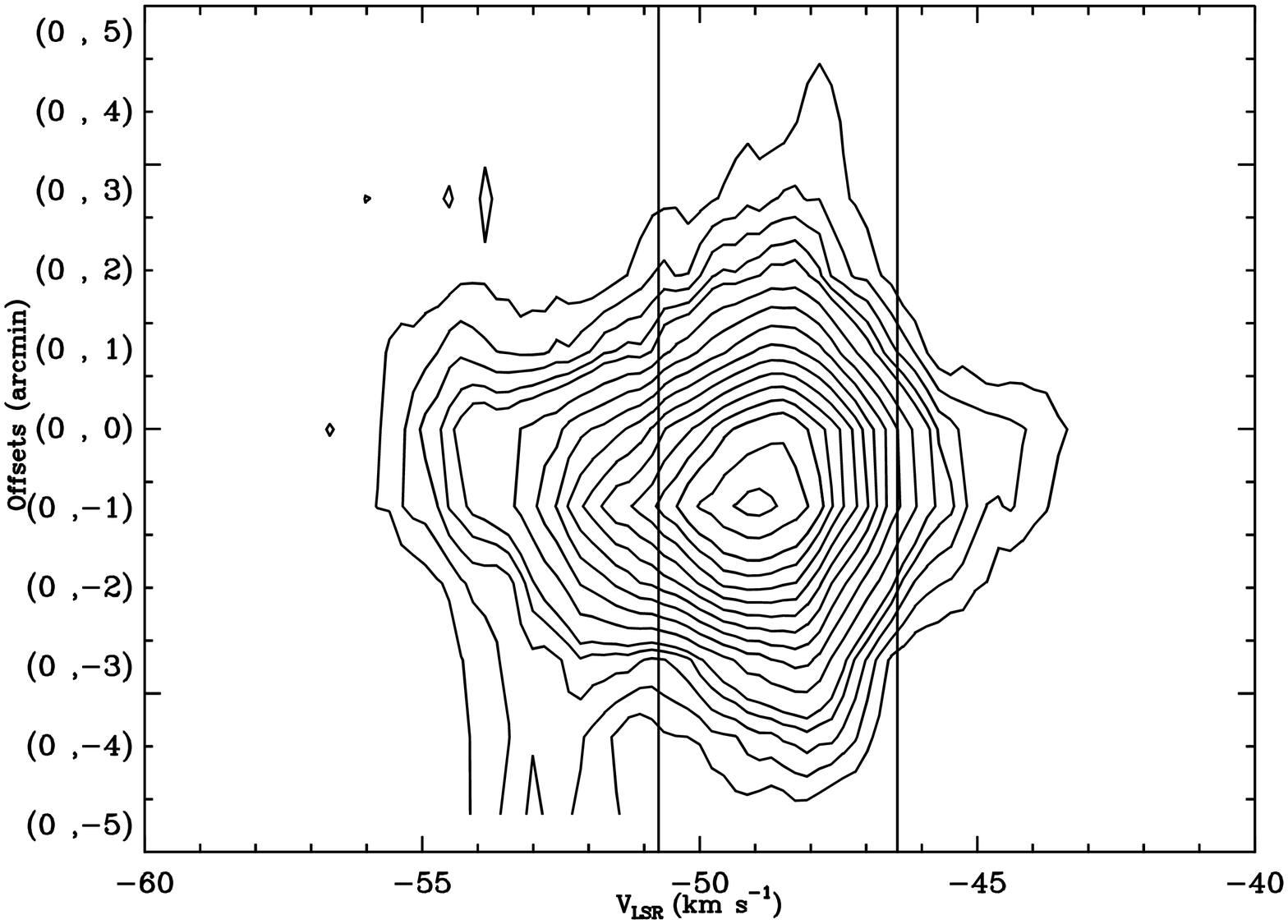}
  \end{minipage}%
  \caption{Left panel:  outflow contours of $\rm CO\, (J=2-1)$, the integrated ranges in blue wing and red wing
  are $-55.78 \rm \,km\cdot s^{-1} \sim -50.80\,\rm  km\cdot s^{-1}$, $-46.05\,\rm km\cdot s^{-1} \sim -43.41\,\rm km\cdot s^{-1}$, respectively.
  The contour levers are $30\% \sim 90\%$ of each wing's peak value. The star is IRAS22475+5939, the triangle is the
   $\rm H_2O$ maser, and the square is the 2MASS source
   22492900+59545600.
  Right panel: P-V diagram. The contour levers are $0.7 \sim 3$ by $0.6\,\rm K$, $3\sim 12$ by $1.0\,\rm K $,
  $12\sim 20$ by $1.5\,\rm K$. The vertical lines indicate the beginning of the blue and red wings, respectively. }
  \label{Fig 2}
  \end{figure}
\begin{table}[h]
\bc
 \small
 \centering

\begin{minipage}[]{70mm}
\caption[]{Physical parameters of the outflow} \label{Tab
2}\end{minipage} \small
 \begin{tabular}{cccccc}
 \hline\noalign{\smallskip}
       $\rm Name$ &$ \rm Wing$ &  $\rm N(H_2) $            & $\rm M$        & $\rm t_d $            &$\rm\dot{M} $         \\
       $\rm IRAS$ &      & $\rm(\times10^{20} cm^{-2})$ &$\rm(M_\odot)$ & $\rm(\times10^5 yr)$ &$\rm(\times10^{-6}M_\odot yr^{-1})$
         \\   \hline\noalign{\smallskip}
$22475+5939$  & $\rm blue$ &    $ 1.00 $  & $278.79$ & $5.37$ & $3.86$ \\
            &$\rm red  $&  $0.15 $  & $26.63 $& $3.25$ &$ 0.80$ \\
     \noalign{\smallskip}\hline
       \end{tabular}

\begin{tabular}{cccccc}
 \hline\noalign{\smallskip}
       $\rm Name$ &$ \rm Wing$         &$\rm F$ &$\rm P$                 &$\rm E $                 & $\rm L_{Mech}$ \\
       $\rm IRAS$ & &$\rm(M_\odot km s^{-1}\cdot yr^{-1})$ &$\rm(M_\odot km s^{-1})$ &$\rm(\times10^{46} erg)$ & $\rm (L_\odot)$  \\   \hline\noalign{\smallskip}
$22475+5939$  & $\rm blue$  & $1.93\times10^{-3}$ &$1028.74$ &$7.59$&$1.17$\\
            &$\rm red  $ &$3.99\times10^{-4}$ &$129.699 $&$1.26$&$1.23$\\
    \noalign{\smallskip} \hline
       \end{tabular}
       \ec
       \end{table}
We show the outflow and its corresponding P-V diagram in Figure 2.
The integrated ranges of the blue wing and the red wing are
determined by the P-V diagram and gaussian fit. We find they overlap
with each other, which is likely they are in the direction of the
sight or the resolution of the telescope is too low to identify the
direction of the outflows. And IRAS22475+5939 as well as its
associated $\rm H_2O$ maser are almost in the center of the wings.
Maybe IRAS22475+5939 is the driving source of the outflow. The
physical parameters are presented in Table 2. Assuming it is local
thermodynamic equilibrium (short for LTE latter) and $\rm CO$ in the
outflow is optically thin, we can calculate the column density of
the outflow using the below formula (Scoville et al. 1986) under the
assumption of  $ {\rm N(CO)/N(H_2)}\approx 10^{-4 }\ (\rm Dickman
\,et \,al. 1978)$:

\begin{equation} \label{eq:1}
\rm {N=10^{5}\times \frac{3k^2}{4h\pi^3\mu^2\nu^2} \exp(\frac{h\nu
J}{2kT_{ex}})
\frac{T_{ex}+{h\nu/{6k(J+1)}}}{\exp(-h\nu/{kT_{ex})}}\times \int{\frac{\tau}{1-e^{-\tau}}T_{mb}}dv }\\
 \end{equation}
where $\rm T_{ex}$ is the excited temperature, which is got from the
equ(9). $\rm J$ is the lower lever of the transition. $\rm
\mu=0.112\,D$.

The other physical parameters coming from the below formulas (Xu \&
Wang 2010) :
\begin{equation} \label{eq:2}
\rm M={um_{H_2}SN(H_2)}/{2\times 10^{33}}
\end{equation}
\begin{equation} \label{eq:3}
\rm P=MV
\end{equation}
\begin{equation} \label{eq:4}
\rm E=MV^2
\end{equation}
\begin{equation} \label{eq:5}
\rm t_d={R}/{V}
\end{equation}
\begin{equation} \label{eq:6}
\rm F={P}/{t_d}
\end{equation}
\begin{equation} \label{eq:7}
\rm \dot{M}={P}/{t_d v_w}
\end{equation}
\begin{equation} \label{eq:8}
\rm L_{Mech}={E}/{t_d}
\end{equation}
where $\rm M$ is the outflow's mass, $\rm t_d$, $\rm P$, $\rm E$,
$\rm V$, $\rm R$, $\rm F$, $\rm L_{Mech}$ are the dynamic time, the
momentum, the energy, the mean velocity of the gas relative to the
$\rm V_{LSR}$, the size of the wings, the driving force, the
mechanical luminosity. The mean atomic weight of the gas $\rm
u=1.36$, $\rm S$ is the area of the outflow. From these parameters,
we can see the outflow is massive and energetic than the low-mass
stars, so we further validate that it is possible to have massive
stars forming in this region. At the same time, through the outflow
and its parameters, we find the blue outflow is far larger than the
red one, and both are elongated from east to west, very similar to
the ${}^{13}\rm CO\, J=3-2$ molecular core that is presented latter.
The explanation is that the surrounding gas might mix with the blue
wing. But we can not make sure because of the low resolution of the
telescope.

\subsection{The molecular core}
\begin{figure}[t]
  \begin{minipage}[t]{0.495\linewidth}
  \centering
   \includegraphics[width=60mm,height=55mm]{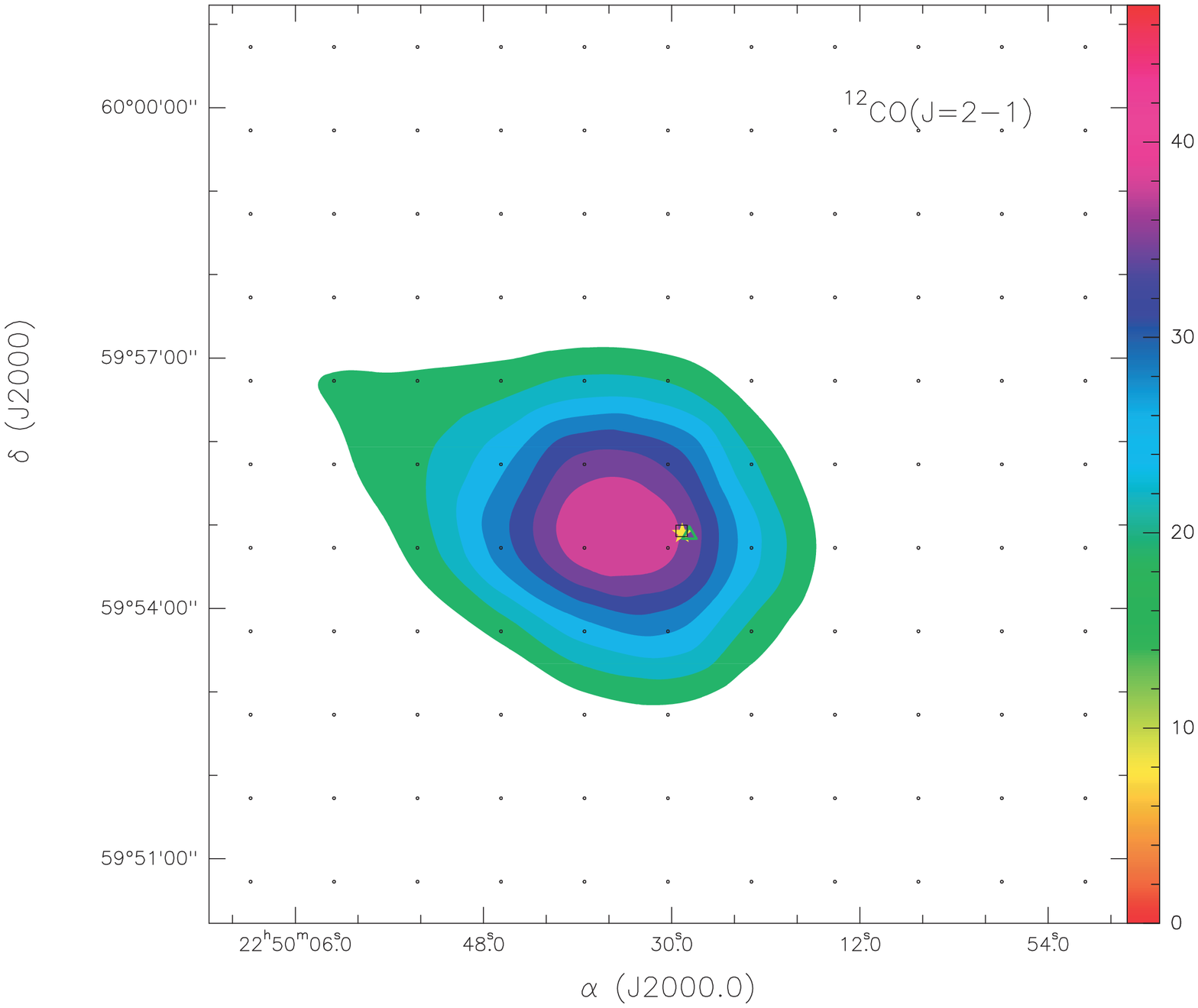}
  \end{minipage}%
  \begin{minipage}[t]{0.495\textwidth}
  \centering
   \includegraphics[width=60mm,height=55mm]{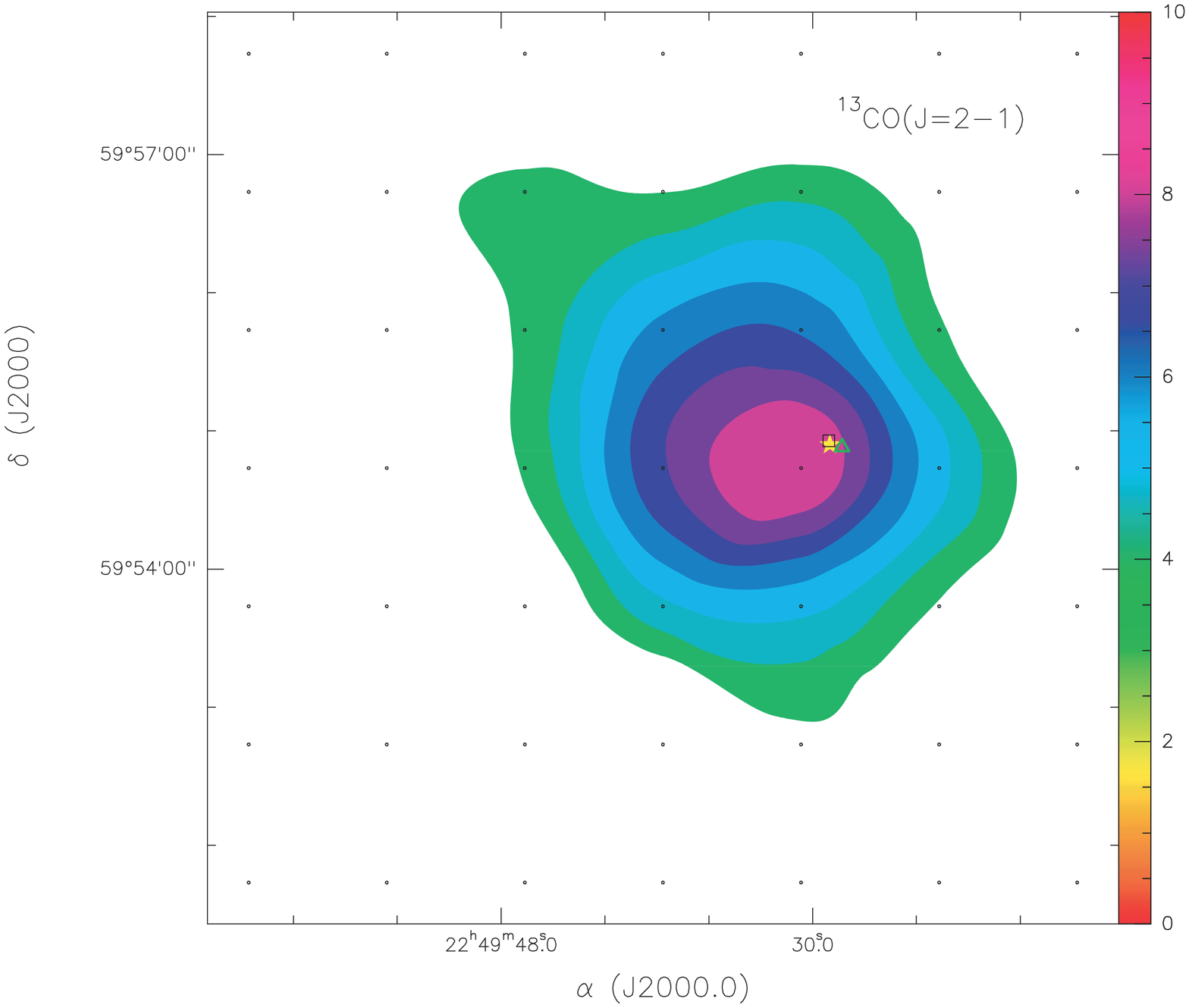}
  \end{minipage}%
\end{figure}
\begin{figure}[t]
  \begin{minipage}[t]{0.495\linewidth}
  \centering
  \includegraphics[width=60mm,height=55mm]{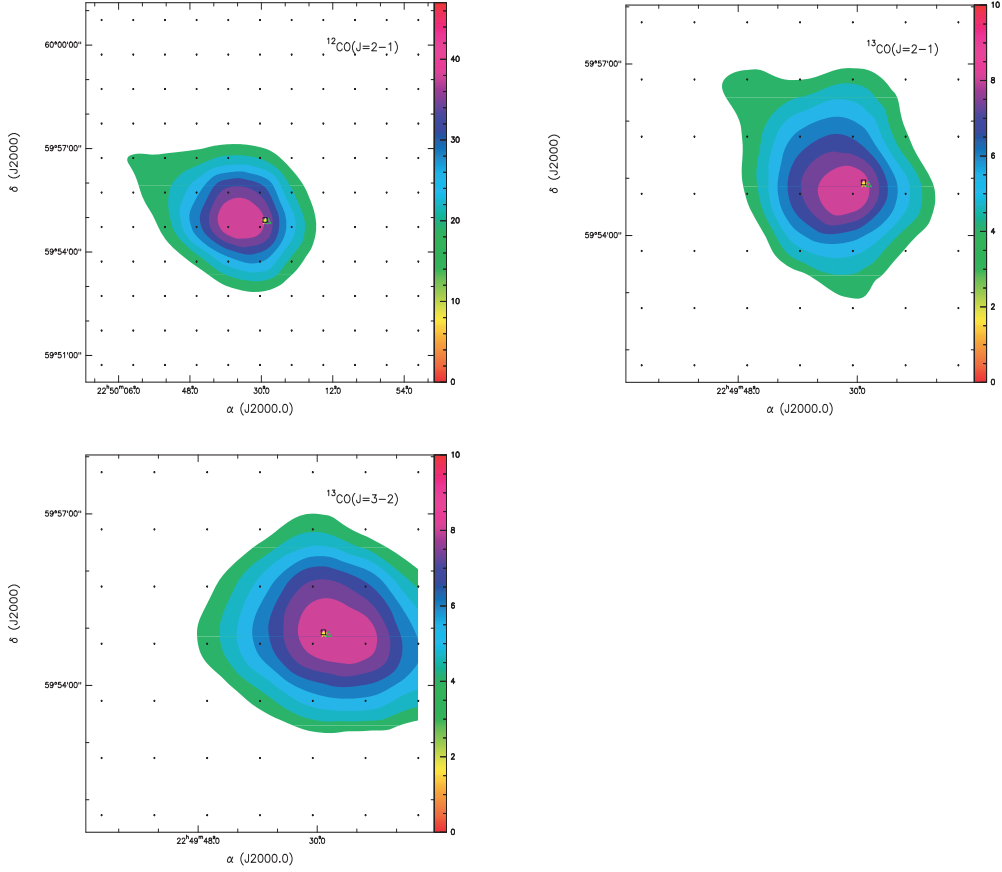}
  \end{minipage}%
  \caption{Integrated intensity diagrams of the core emission. In each map, the integrated range is
  from $-50.80\rm\, km\cdot s^{-1} $to $-46.05\rm\, km\cdot s^{-1}$. The contour levels are $30\%$ to $90\% $ of the peak value.  }
  \label{Fig 3}
  \end{figure}
Figure 3 shows the structure of the molecular core. The integrated
range is determined by the P-V diagram and the gaussian fit due to
the relatively large noise. It is more accurate than one determined
by each of them. In the diagrams, we find the core structure from
${}^{13}\rm CO\, J=3-2$ is elongated from east to west, but almost
round from the transition $\rm CO \,J=2-1$, ${}^{13}\rm CO\,J=2-1$.
IRAS 22475+5939 and $\rm H_2O$ maser coincide with the peak of
cores. We also calculate the core's physical parameters by the
expressions (Garden et al. 1991) under the assumption of LTE:
\begin{equation} \label{eq:9}
\rm
T_{ex}(CO)=\frac{h\nu}{k}\big(\ln({1+\frac{h\nu}{k}[\frac{T_{mb}}{f}+\frac{{h\nu}/{k}}{\exp({h\nu}/{kT_{bg}})-1}]^{-1}})\big)^{-1}
\end{equation}
\begin{equation} \label{eq:10}
\rm\tau({}^{13}CO)=-\ln\big\{1-\frac{kT_{mb}}{h\nu}[\frac{1}{\exp({h\nu}/{kT_{ex}})-1}-\frac{1}{\exp({h\nu}/{kT_{bg}})-1}]^{-1}\big\}
\end{equation}
where $\rm T_{bg}=2.732\, K $ is the temperature of the cosmic
background radiation, f is the beam filling factor, here we assuming
 $f=1$. Assuming
${\tau(\rm CO)}/{\tau({}^{13}\rm CO)}={[\rm CO]}/{[{}^{13}\rm
CO]}=89\, (\rm Lang\, et \,al.1980)$. We can use another method to
calculate the optical depths:
\begin{equation}
\rm\frac{T_{mb}(CO)}{T_{mb}({}^{13}CO)}=\frac{1-\exp[-\tau(CO)]}{1-\exp[-\tau({}^{13}CO)]}
\end{equation}
The two calculated values are summarized in Table 3. Comparing with
them, we find they are the same in the error scale. Therefore we can
draw a conclusion that our assumptions are right and the abundance
ratio $ {[\rm CO]}/{[{}^{13}\rm CO]} $  in this region is almost the
same as in our solar system. The core's physical parameters are in
Table 4. From the figures and tables, we can know that the core is
isolated and massive, which can provide the environment for the
high-mass star formation.

\begin{table}[h!!!]
\bc
 \centering

\begin{minipage}[]{90mm}
 \caption[]{Calculated results of the optical depth by
two methods}
 \label{Tab 3} \end{minipage} \\
 \tabcolsep 8mm
\begin{tabular}{cccccc}
\hline
$\rm Name$ &  $\rm \tau_1$   &$\rm \tau _2$        \\
\hline
${}^{13}\rm CO(J=2-1)$ & $0.28$     & $0.28 $         \\
\hline
        \end{tabular}
       \ec
       \tablecomments{0.6\textwidth}{$\tau_1$ is derived from the fist method,
       $\tau_2 $ is from the second method.}
       \end{table}

\begin{table}[h!!!]
\bc
 \small
 \centering

\begin{minipage}[]{70mm}
 \caption[]{Physical parameters of the core}
 \label{Tab 4}\end{minipage}
\begin{tabular}{cccccc}
 \hline\noalign{\smallskip}
$\rm Name$              &$\rm T_{ex}$   &$\rm \tau $  &$N(\rm CO J=2-1) $         &$N(\rm H_2) $               &$M $                       \\
                   & $\rm(K)$    &       &$\rm(\times 10^{18} cm^{-2})$ &$\rm (\times 10^{22} cm^{-2})$ & $\rm(\times 10^3 M_\odot)$   \\ \hline
$\rm IRAS 22475+5939 $   & $17.35 $ & $24.92$ &$ 1.03$                & $1.03 $                & $3.56 $                    \\
\noalign{\smallskip}\hline
       \end{tabular}
       \ec
       \end{table}

\subsection{The intensity ratio $R_{I_{32/21}}$ }
\begin{figure}[h!!!]
  \centering
  \includegraphics[width=60mm,height=55mm]{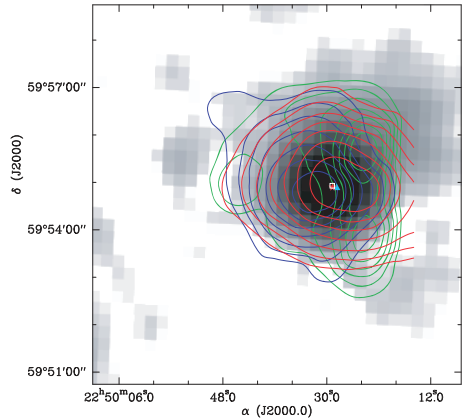}\\
  \caption{Grayscale shows the $1.4$ GHZ NVSS image. The intensity ratio $\rm R_{I_{32/21}}$
  is the green contour, the blue and red contours are the integrate intensity maps of ${}^{13}\rm CO\,(J=2-1)$,
   ${}^{13}\rm CO \,(J=3-2)$, respectively. The star is IRAS22475+5939, the triangle is the
   $\rm H_2O$ maser, and the square is the 2MASS source 22492900+59545600.}
  \label{fig 4}
  \end{figure}
Figure 4 gives the contours of ${}^{13}\rm CO$ and the intensity
ratio $\rm R_{I_{32/21}}$ overlapping on the $1.4$ GHZ NVSS
grayscale diagram which can trace the structure of the radio source,
here we can identify the radio source is a HII region (Lockman et
al. 1989). The intensity ratio indicates the gas temperature
distribution in the region (Qin et al. 2008). From Fig 4, we can
find the morphology of the intensity ratio map is similar to a
triangle and has three peaks. The peaks are not in the center of the
molecular cloud core, the biggest peak is in the northwest, about
$2'$ away from the center. The value of it is $\sim $2.12 $>1$,
which is bigger than the value ($<1$) detected in Cepheus B (Beuther
et al. 2000). Maybe it is related with HII regions (Wilson et al.
1997). The second peak and the third peak are in the direction of
southwest and east, respectively. All of three are at the edge of
the HII region. This shows the temperature at the edge is higher
than the center. Maybe the hot gas of the molecular cloud was pushed
away to the edge and heated by the HII region. This explanation can
be supported by the distribution of the HII region and the molecular
core. Additionally, we find $\rm {}^{13}CO$ higher J can trace the
warmer region. And the center of the molecular core coincides with
that of the HII region.

\subsection{polycyclic aromatic hydrocarbons (PAHs)}
\begin{figure}[t]
  \begin{minipage}[t]{0.495\linewidth}
  \centering
   \includegraphics[width=60mm,height=55mm]{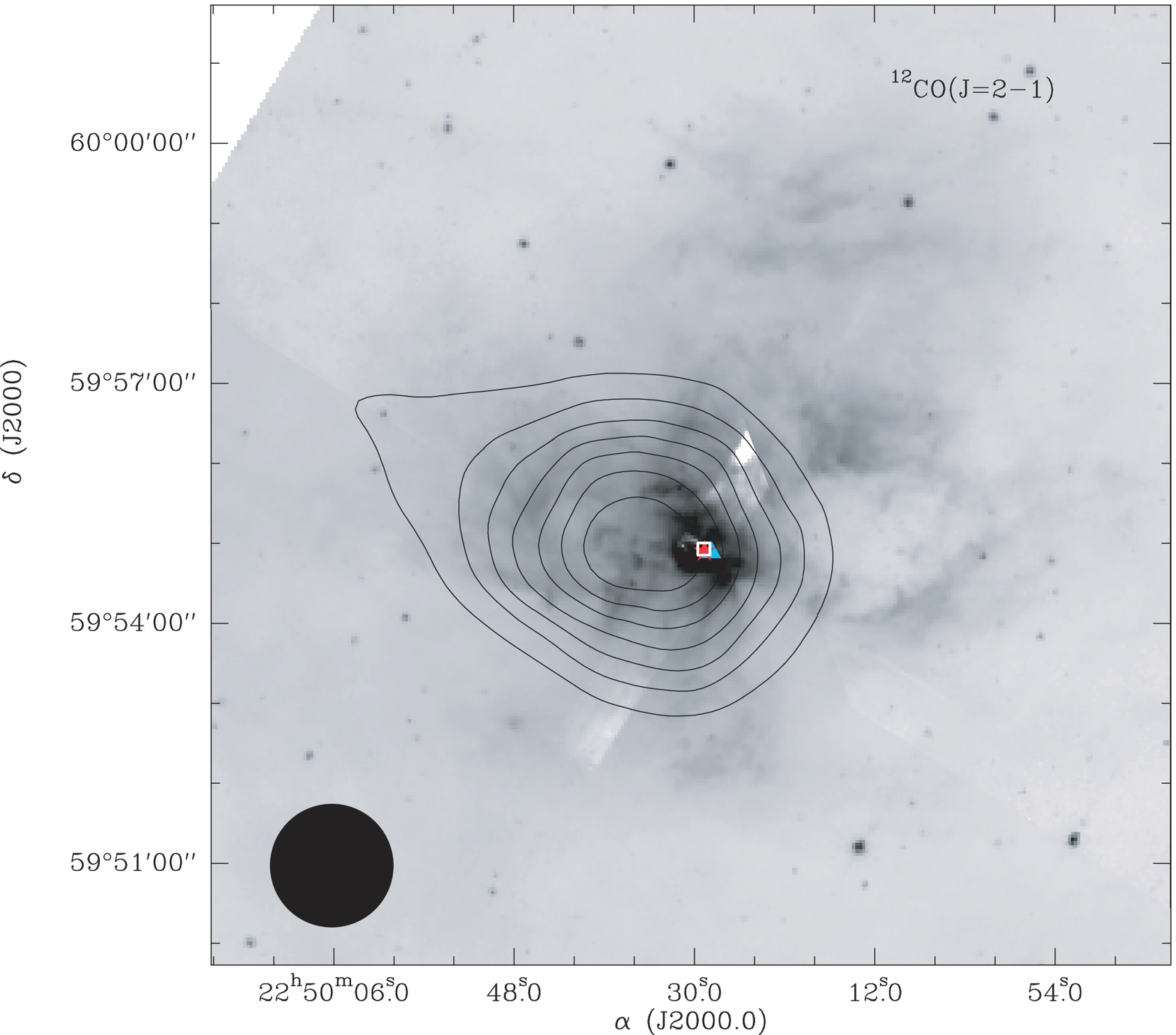}
  \end{minipage}%
  \begin{minipage}[t]{0.495\textwidth}
  \centering
   \includegraphics[width=60mm,height=55mm]{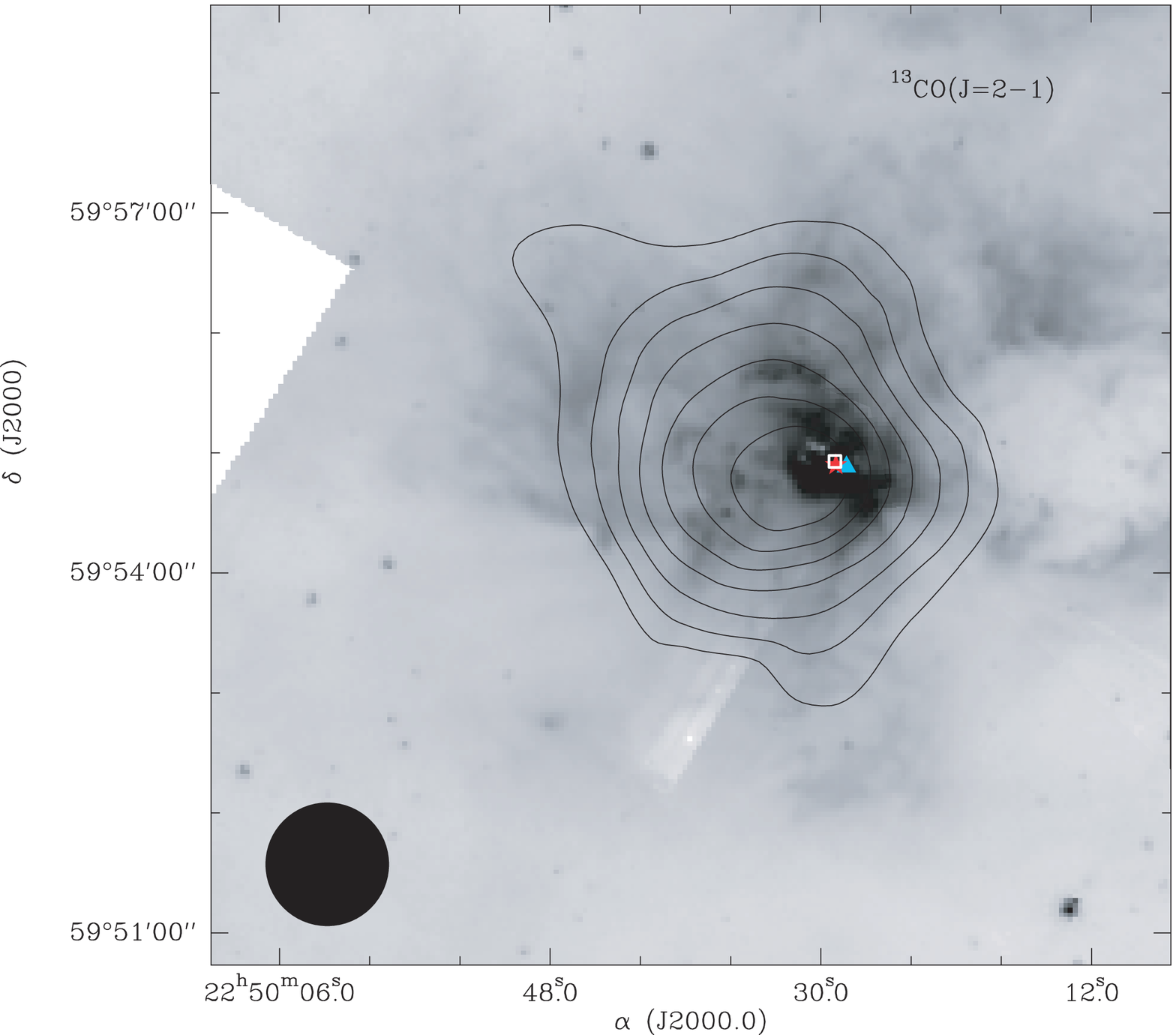}
  \end{minipage}%
\end{figure}
\begin{figure}[t]
  \begin{minipage}[t]{0.495\linewidth}
  \centering
   \includegraphics[width=60mm,height=55mm]{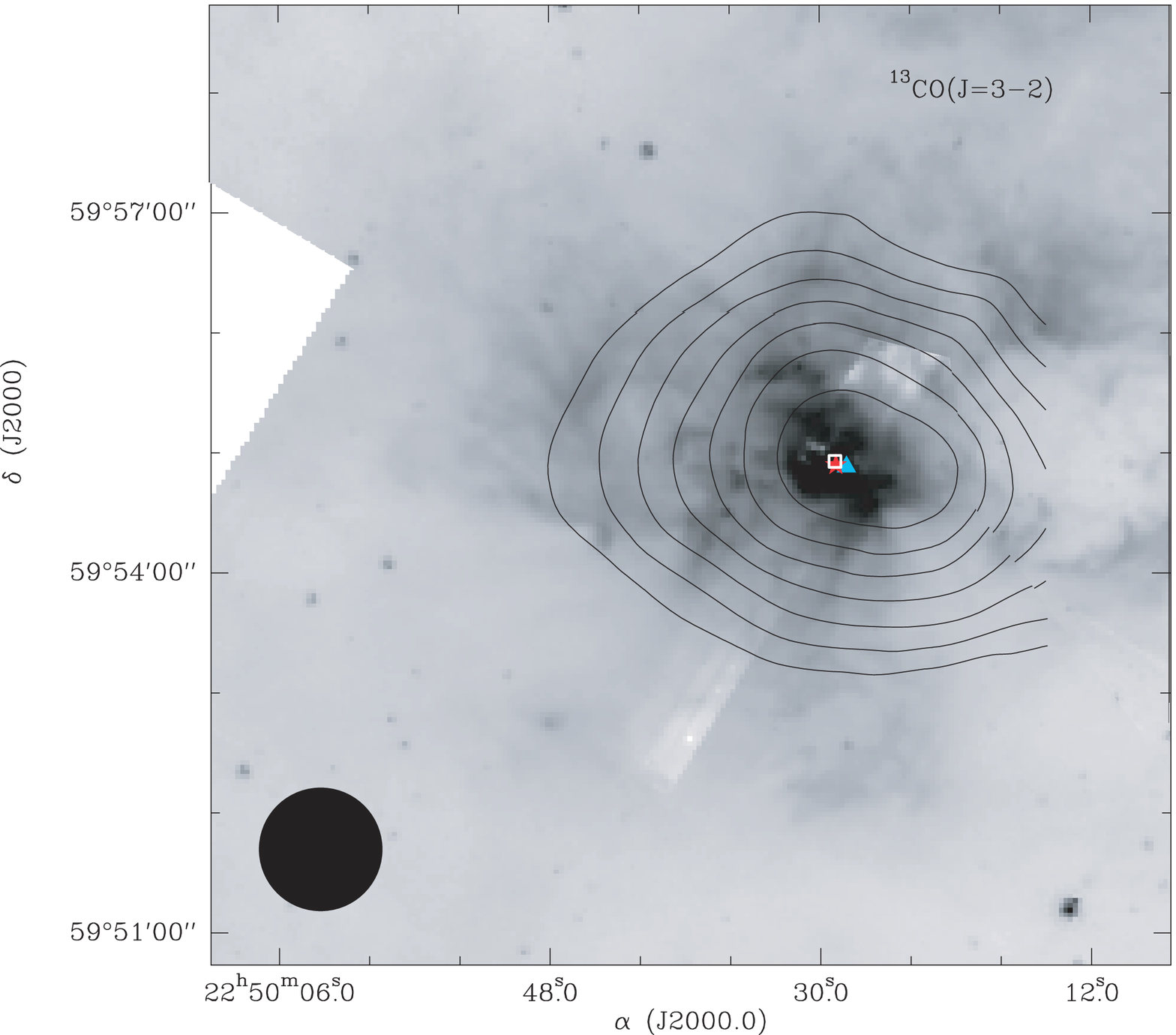}
  \end{minipage}%
  \begin{minipage}[t]{0.495\textwidth}
  \centering
   \includegraphics[width=60mm,height=55mm]{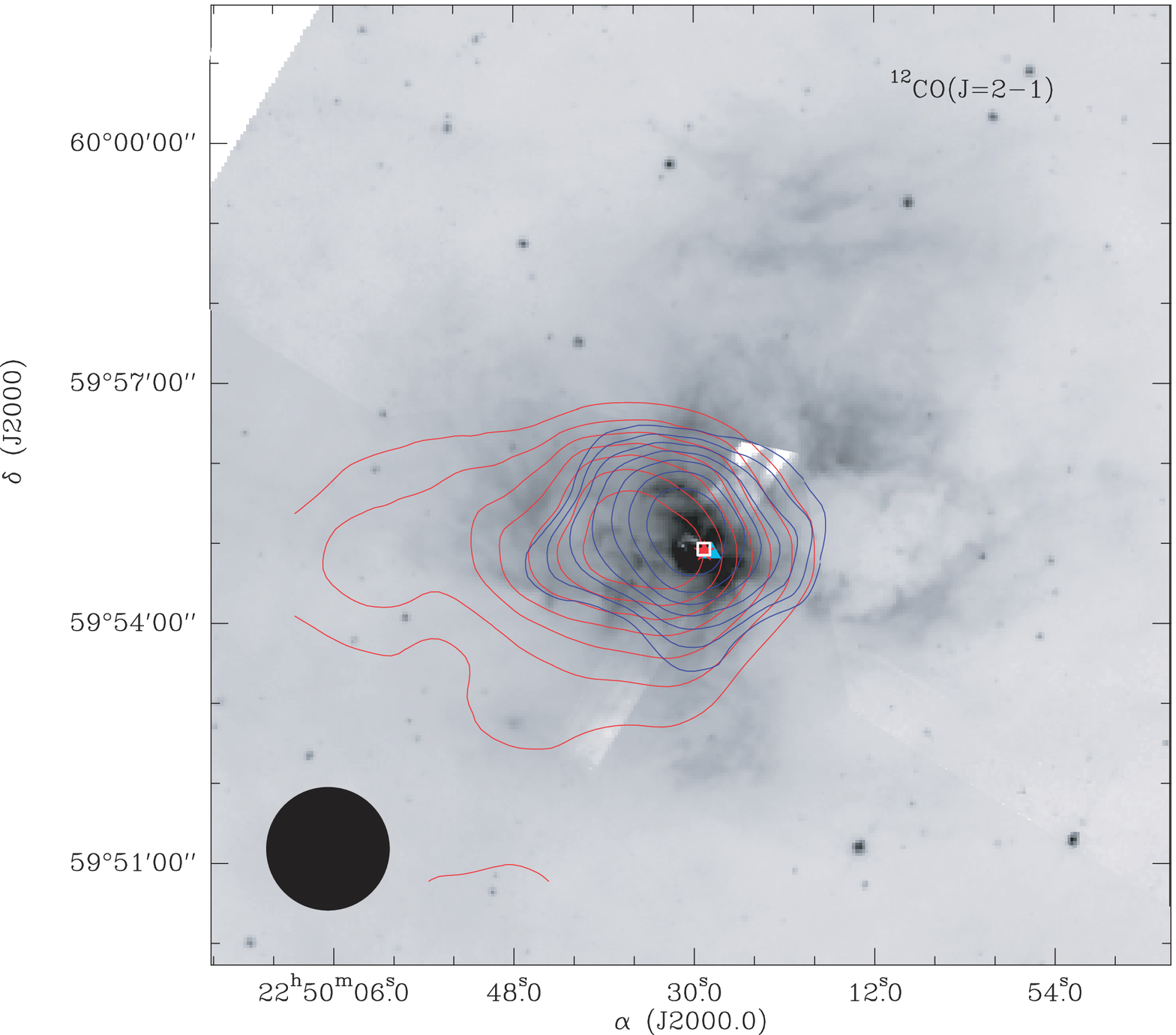}
  \end{minipage}%
  \caption{The molecular cores from the transitions $\rm CO\, (J=2-1)$, ${}^{13}\rm CO\, (J=2-1)$, ${}^{13}\rm CO
  \,(J=3-2)$
  (top left, top right, bottom left) and the outflow of $\rm CO \, (J=2-1)$ emission (bottom right)
   overlaid on the grayscale Spitzer $ 8 \,um$ emission. The synthesized beam size ($\sim 1'$) is
   shown in the lower left-hand corner of the each panel of KOSMA 3m telescope. }
  \label{Fig 5}
  \end{figure}

  \begin{figure}
  \begin{minipage}[t]{0.495\linewidth}
  \centering
   \includegraphics[width=60mm,height=55mm]{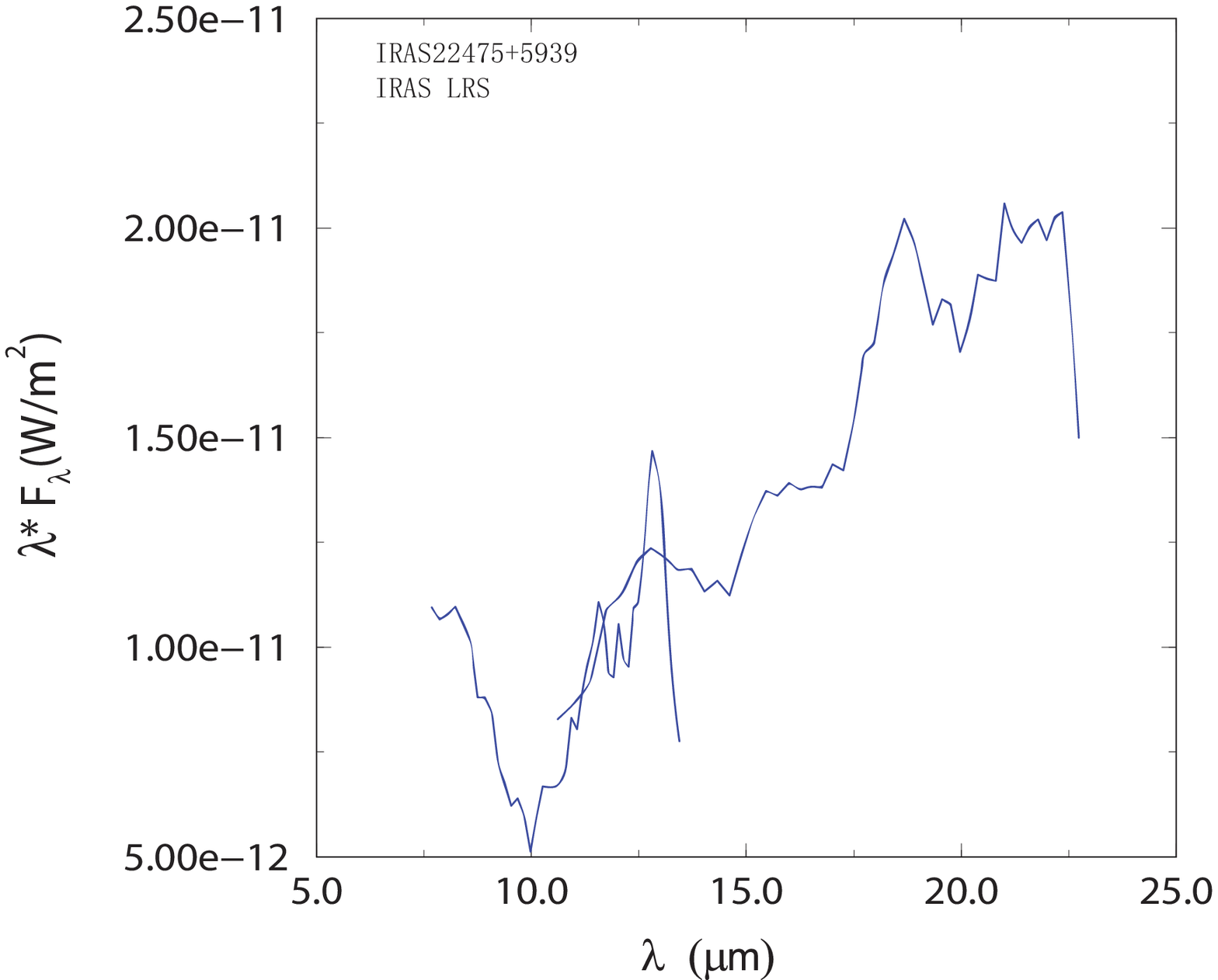}
  \end{minipage}%
  \begin{minipage}[t]{0.495\textwidth}
  \centering
   \includegraphics[width=60mm,height=55mm]{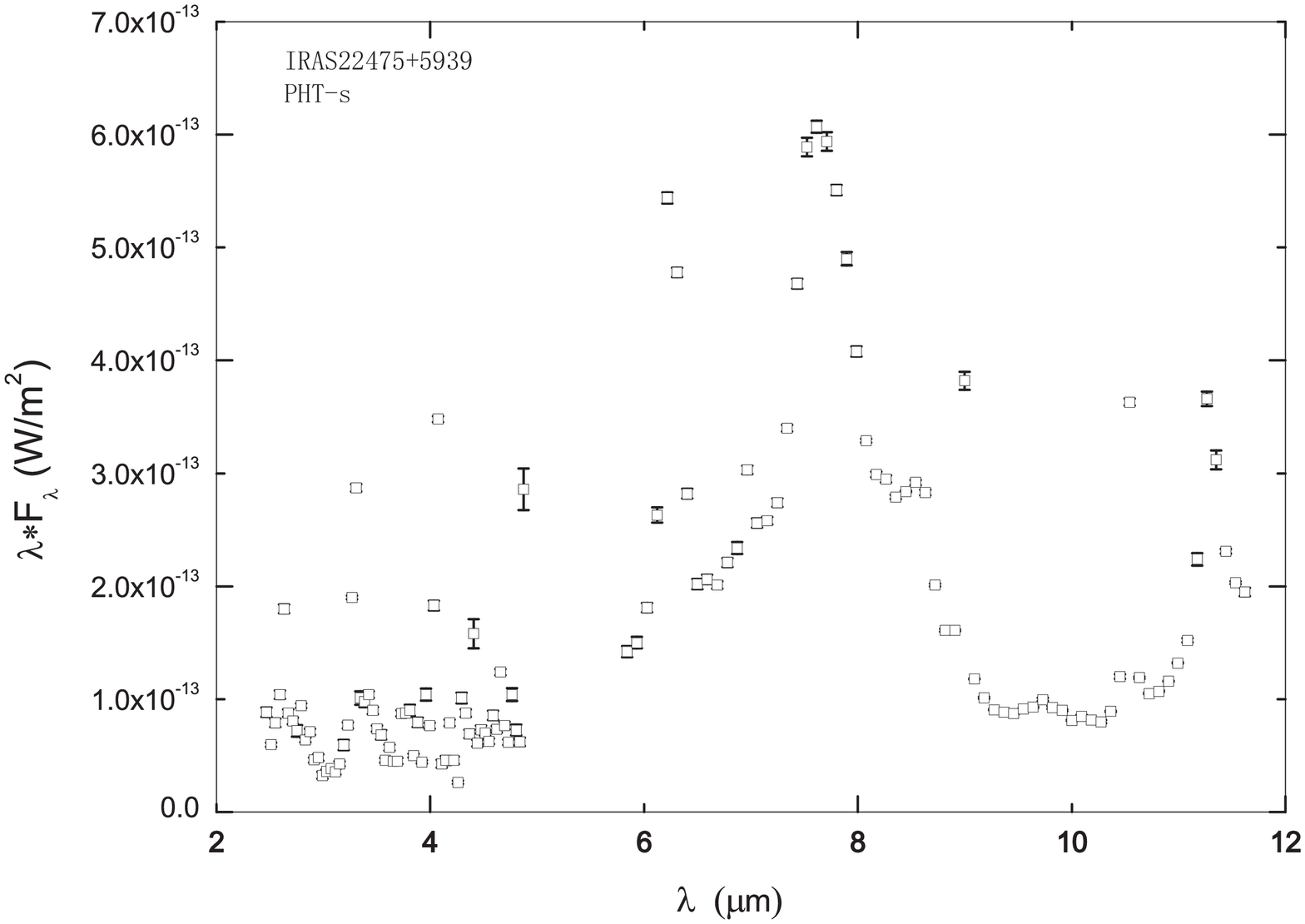}
  \end{minipage}%
  \caption{Left panel: the IRAS-LRS spectrum coming from LRS database at the university of Calgary,
  consistent of two wavelength band:$7.7-13.4 \,\mu m$, $11.0-22.6 \,\mu m$. Right panel: the PHT-s spectrum lying in the $2-12\,\mu m$ range.   }
  \label{Fig 6}
  \end{figure}
Fig 5 shows the superposition diagrams of the $\rm CO $ isotopes'
contours and $\rm CO $ outflow on the Spitzer $8 \,\mu m$ emission.
The Spitzer IRAC $8 \,\mu m$ emission is primarily due to the $7.7
\,um$ and $8.6 \,\mu m$ PAHs features. The grayscale diagram shows
the distribution of PAHs in the HII region. We can see that the
strongest PAHs emission coincides with IRAS22475+5939 source, which
shows the PAHs emission probably is excited by the UV radiation from
IRAS22475+5939.

Fig 6 shows the LRS spectrum and the PHT-s spectrum, spanning the
wavelength $7.7-23\,\mu m$ and $2-12\,\mu m$, respectively. They are
dominated by PAHs emission features at the so-called unidentified
infrared bands ($3.3, 6.2, 7.7, 8.6, 11.2 \,\mu m$) and weaker bands
\footnote{This includes bands at $3.4, 3.5, 5.25, 5.75, 6.0, 6.6,
6.9, 7.2-7.4, 8.2, 10.5, 10.8,  11.0, 12.0, 13.5, 14.2, 15.8, 16.4,\\
16.6, 17.0, 17.4, 17.8, 19.0 \,\mu m$}(Peeters 2011). The PHT-s
spectrum shows the narrow and strongest emission at $7.6 \,\mu m$ at
the wavelength range $2-12\,\mu m$ and an absorption at $9.7 \,\mu
m$, but the strongest emission for the LRS is at $12.0 \,\mu m$ at
band 1. It is more believable for the PHT-s due to its higher
resolution than the results of the LRS. And a broad emission plateau
at $16.4-17.4 \,\mu m$ maybe resolute from the low resolution or the
diverse PAH family. Muizon (1990) pointed out that the prominent
emissions at $12.8, 15.55, 18.7 \,\mu m$ are [NeII], [NeIII], [SIII]
lines,respectively.

\subsection{The spectral energy distribution}
\begin{figure}
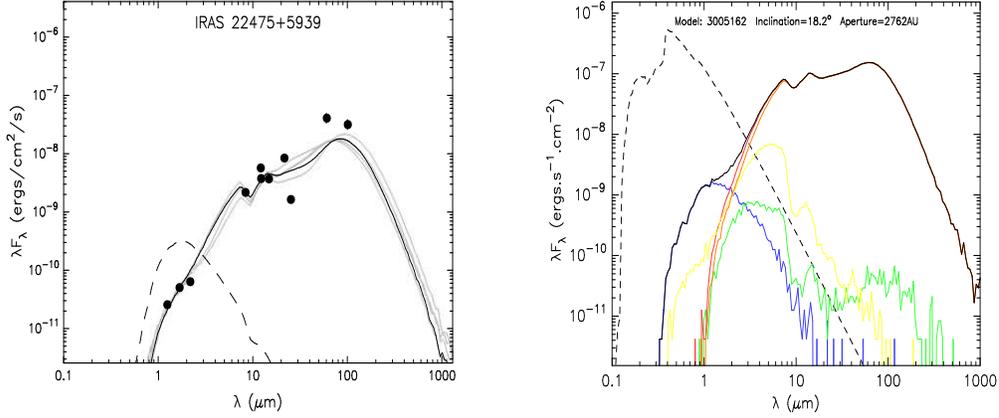

  \begin{minipage}[t]{0.495\linewidth}
  \centering
   \includegraphics[width=60mm,height=55mm]{IRAS22475+5939f19.eps}
  \end{minipage}%
  \begin{minipage}[t]{0.495\textwidth}
  \centering
   \includegraphics[width=60mm,height=55mm]{IRAS22475+5939f20.eps}
  \end{minipage}%
  \caption{Left panel:the SED of the IRAS 22475+5939 source. The filled circles show the data from 2MASS (JHK), MSX (ACDE), IRAS ($12,\,25,\,60,\,100 \,um$).
  The continuous line shows the best fit model and the grey lines show subsequent good fits for ($\chi^2-\chi_{bestfit}^2$ ) per data point $< 3$.
  The dashed line corresponds to the stellar photosphere for the central source of the best fitting model, as it would look in the absence of the
  circumstellar (but including interstellar extinction). Right panel: various emission components making up this model, details in the text.
  Suffice to remark that the circumstellar disc (green curve) is important for wavelengths between $\rm 1 \,and \,10 \,\mu m$.}
  \label{Fig 7}
  \end{figure}

\begin{table}[h!!!]
\bc
 \small
 \centering
  \begin{minipage}[]{50mm}
 \caption[]{Results from SED fitting}
 \label{Tab 5}\end{minipage}
 \tabcolsep 1.5mm
\begin{tabular}{cccccccccccc}
 \hline\noalign{\smallskip}
&$\rm Parameter$    &$\rm SM$ &$\rm  age $  &$\rm T_{EFF} $  &$\rm DM $   &$\rm Disk \,accretion\, rate$  &$\rm Luminosity$ &$\rm A_v$           \\
  &                 & $\rm(M_\odot)$      &$\rm(yr)$             & $\rm(K)$               & $\rm(M_\odot)$ &$\rm(M_\odot yr^{-1})$  &$\rm L_\odot$ & \\ \hline
&$\rm Best\, fit\, values $   & $15.34 $  & $1.54\times 10^4$    & $9995.8 $                & $0.13 $        & $1.56\times 10^{-5} $  &$1.54\times 10^4$  &$4.53$          \\
\noalign{\smallskip}\hline
       \end{tabular}
       \ec
       \tablecomments{0.85\textwidth}{$\rm SM$ is short for stellar mass, $\rm DM $ is for disk mass,
        the luminosity is the total luminosity of the star, $\rm A_v$ is the foreground extinction.
        The best fit values are derived the average values for the best top $10$ models.   }
       \end{table}

 The 2MASS, MSX, and IRAS data are used to construct the SED of the
 IRAS22475+5939 source. Color correction is applied to the IRAS
 data using the correction factors given in the point source
 catalogue (Beichman et al. 1988).
 The SED fitting tool of Robitaille et al. (2007) is available
 on-line to model the SED. The SED plot is showed in Fig 7 and its
 fitting parameters in Table 5. We derived a mass, luminosity and
 temperature of $15.34 \rm\,M_\odot $, $1.54\times 10^4 \,\rm
 L_\odot$, $9995.8 \rm\, K$ respectively from the fit. And the
 average foreground extinction is $\sim 4.53$. From the SED and the
 active accreting mass, IRAS22475+5939 is probably to be a class I protostar.

 We have also indicated the differing component of flux which make up
 the model YSO spectrum (right panel of Fig 7). In this case, the
 total flux is indicated as black, the stellar flux as blue, the
 stellar photospheric flux is showed in the dashed line (this is
 the flux prior to reddening by circumstellar dust), the disc flux
 as green, the scattered flux as yellow, the envelope flux as red
 and the thermal flux as orange. Unless otherwise stated, the
 results include the effects of circumstellar extinction, but not of
 IS extinction. The also assume a representative distance of
 $1\,\rm kpc$.

 It is apparent, from the latter modeling, that the central star and
 disc components of emission are responsible for the MIR/NIR
 emission, but envelope fluxes are of most importance at $\lambda > 10\,\mu
 m$.

\subsection{Discussion}
The morphology of the cores are firstly showed from the lines $ \rm
CO\, J=2-1$, ${}^{13}\rm CO \,J=2-1$, ${}^{13}\rm CO\, J=3-2$, as
well as the corresponding parameters. Comparing with the results in
$ \rm CO,\,{}^{13}CO,\,{}^{18}CO\, J=1-0$ (Guan et al. 2008), the
$\rm V_{LSR}$ and the derived physic parameters are litter
difference. And the structure of ${}^{13}\rm CO \,J=1-0$ detected by
them is similar to ${}^{13}\rm CO \,J=2-1$. The outflow traced by $
\rm CO\, J=2-1$ is more extended than that of $ \rm CO\, J=3-2$
(Jiang et al. 2001). Yang(1998) only analyzed the IRAS data and
obtained that IRAS22475+5939 may be a massive YSO with high
luminosity. Due to the SED in infrared bands, we suggest that
IRAS22475+5939 is likely to be a class I YSO with $\rm mass\, 15.34
\,M_\odot, lumnosity\,1.54\times10^4\,L_\odot, age
\,1.54\times10^4\,yr, T_{EFF}\,9995.8\,K$. Its foreground extinction
is about $4.53\rm \,mag$. Muizon (1990) suggests the PAH
characteristics at $7.7,8.6,11.3 \,\mu m$, silicate absorption
features at $9.7\,\mu m$ and the prominent emissions at $\rm
12.8\,\mu m([ NeII]),\, 15.55\,\mu m([NeIII]),\,18.7\,\mu m ([SIII])
 $ according to the IRAS-LRS spectrum. But we combine the
IRAS-LRS spectrum with the PHT-s spectrum, finding the PAHs
emissions are showed at all the PAH bands, especially at the main
PAH bands ($3.3, 6.2, 7.7, 8.6,11.2 \, \mu m$) and $16.4-17.4 \,\mu
m$. This indicates the diverse PAH family in the molecular cloud.
The HII region is likely to be excited by IR1-an O7V-O7.5V star
(Eiroa et al.1981). Its position almost coincides with
IRAS22475+5939. Therefore, the star has the chance to drive the
outflow. It is necessary to have high resolution observation to
identify their relationship.
\section{Conclusion}
We observed the IRAS22475+5939 source using three spectral lines $
\rm CO\, J=2-1$, ${}^{13}\rm CO \,J=2-1$, ${}^{13}\rm CO\, J=3-2$.
The $\rm V_{SLR}$ of them are almost the same. The massive and
energetic outflow is further verified by us, showing that this
region is a high-mass star formation region. The integrated
intensity maps of the cores tell us the molecular cloud is isolated.
And the core's mass is larger than the low-mass cores. The intensity
ratio $\rm R_{I_{32/21}}$ map indicates the gas temperature varies
at different positions and the maximum value is bigger than 1, which
is larger than the regions without HII regions (Wilson et al. 1997).
And the peaks of the intensity ratio are at the edge of the HII
region. Maybe the HII region blows away the hot molecular gas from
the center to the edge and heats the gas. This causes the
temperature is higher at the edge than the center of the molecular
core. And ${}^{13}\rm CO$ higher J transition traces the warm
regions. PAH features are demonstrated in the Spizter IRAS $8\, \mu
m$ emission and the spectra of IRAS-LRS and PHT-s, suggesting a rich
PAH molecules in the cloud. IRAS22475+5939 may be a massive and
luminous class I protostar with a active disc. But considering the
position of the excited star of the HII region, the excited star is
able to drive the outflow. Maybe it is the driving source of the
outflow. The high resolution observations is needed for better
results.

  Acknowledgements:We would like to thank Drs.Sheng-Li Qin and Martin Miller
  for the data acquisition.
  This research has made use of the data products from
  the $1.4$ GHZ NRAO VLA Sky Survey (NVSS), and also used the
  NASA/IPAC Infrared Science Archive, which is operated by the Jet
  Propulsion Laboratory, California Institute of Technology, under
  contract with the National Aeronautics and Space Administration.
  IRAS, MSX, 2MASS photometry are download from the NASA/IPAC
  Infrared Science. We thank SUN Kwok for providing the LRS database
  at the University of Calgary. ISOPHOT data was obtained from the
  ISO Data Archive(IDA), tanks to Lemke, D. for his reducing to the
  PHT-s data.

\label{lastpage}

\end{document}